\def\rfr#1{eq. (\ref{#1})}

\def\virg#1{``#1''}

\def\bb#1#2#3{\bibitem[\protect\citeauthoryear{#1}{#2}]{#3}}
\def\eqi{\begin{equation}}
\def\eqf{\end{equation}}
\def\eqia{\begin{eqnarray}}
\def\eqfa{\end{eqnarray}}
\def\rp#1#2{{#1\over#2}} \def\lb#1{\label{#1}}
\def\bb#1#2#3{\bibitem[\protect\citeauthoryear{#1}{#2}]{#3}}

\def\bds#1{\vec{#1}}

\documentclass{elsart}

\usepackage{natbib}

\usepackage{epsfig}

\usepackage{amssymb}

\begin{document}

\begin{frontmatter}



\title{Gravitomagnetism  and the Earth-Mercury range}


\author{Lorenzo Iorio\thanksref{footnote2}}
\address{Ministero dell'Istruzione, dell'Universit\`{a} e della Ricerca (M.I.U.R.)-Istruzione\\
International Institute for Theoretical Physics and
High Mathematics Einstein-Galilei}
\thanks[footnote2]{Address for correspondence:  Viale Unit\`{a} di Italia 68, 70125, Bari (BA), Italy}
\ead{lorenzo.iorio@libero.it}
\ead[url]{http://digilander.libero.it/lorri/homepage$\_$of$\_$lorenzo$\_$iorio.htm}



\begin{abstract}
We numerically work out the impact of the general relativistic Lense-Thirring effect on the Earth-Mercury range $|\bds\rho|$ caused by the gravitomagnetic field of the rotating Sun. The peak-to peak nominal amplitude of the resulting time-varying signal amounts to $1.75\times 10^1$ m over a temporal interval $\Delta t=2$ yr. Future interplanetary laser ranging facilities should reach a cm-level in ranging to Mercury over comparable timescales; for example, the BepiColombo mission, to be launched in 2014, should reach a $4.5-10$ cm level over $1-8$ yr. We looked also at other Newtonian (solar quadrupole mass moment, ring of the minor asteroids, Ceres, Pallas, Vesta, Trans-Neptunian Objects) and post-Newtonian (gravitoelectric Schwarzschild solar field) dynamical effects on the Earth-Mercury range. They act as sources of systematic errors for the Lense-Thirring signal which, in turn, if not properly modeled, may bias the recovery of some key parameters of such other dynamical features of motion. Their nominal peak-to-peak amplitudes are as large as $4\times 10^5$ m (Schwarzschild), $3\times 10^2$ m (Sun's quadrupole),  $8\times 10^1$ m (Ceres, Pallas, Vesta), $4$ m (ring of minor asteroids), $8\times 10^{-1}$ m (Trans-Neptunian Objects). Their temporal patterns are different with respect to that of the gravitomagnetic signal.
\end{abstract}


\end{frontmatter}

\parindent=0.5 cm

\section{Introduction}
In its slow-motion and weak-field approximation, the Einsteinian General Theory of Relativity (GTR) predicts that a slowly rotating central body of mass $M$ and proper angular momentum $\bds S$ induces two kinds of small perturbations on the otherwise Keplerian  motion of a test particle orbiting it. The largest one is dubbed \virg{gravitoelectric} (GE) \citep{Mash}, and depends only on the mass $M$ of the body which acts as source of the gravitational field. It is responsible of the well-known anomalous secular precession of the perihelion $\omega$ of Mercury of $42.98$ arcsec cty$^{-1}$ in the field of the Sun. There is also a smaller perturbation, known as \virg{gravitomagnetic} (GM) \citep{Mash}, which depends on $\bds S$: it causes the Lense-Thirring (LT) precessions of the node $\Omega$ and pericenter $\omega$ of a test particle \citep{LT} .
More specifically,
the Parameterized Post-Newtonian (PPN) perturbing acceleration $\bds A_{\rm PPN}$ to be added to the Newtonian monopole $\bds A_{\rm Newton}\doteq -GM \hat{r}/r^2$ in the equations of motion
\eqi \rp{d^2 \vec r}{dt^2}= \vec A_{\rm Newton}+\vec A_{\rm PPN}\eqf
is \citep[p. 89, p.95]{Soffel}, \citep[p. 106]{IERS}
\eqi \bds A_{\rm PPN}\doteq-\bds E_g -2\left(\rp{\bds v}{c}\right)\times\bds B_g.\lb{gemeq}\eqf
In it \citep[p. 89, p.95]{Soffel}, \citep[p. 106]{IERS},
\eqi
\begin{array}{lll}
\bds E_g &\doteq & -\rp{GM}{c^2 r^3}\left\{\left[\rp{2(\beta+\gamma)GM}{r}-\gamma v^2\right]\bds r+2(1+\gamma)\left(\bds r\cdot\bds v\right)\bds v\right]\}, \\ \\
\bds B_g &\doteq & -\left(\rp{1+\gamma}{2}\right)\rp{G}{cr^3}\left[\bds S - 3\left(\bds S\cdot{\hat{r}} \right){\hat{r}}\right],\lb{egos}
\end{array}
\eqf
where $\gamma,\beta$ are the usual PPN parameters \citep{Will} which are both 1 in GTR, $c$ denotes the speed of light in vacuum, $G$ is the Newtonian constant of gravitation, and $\bds v$ is the velocity of the test particle moving at distance $r$ from $M$; the unit vector $\hat{r}$ is directed from the central body to the test particle.
In \rfr{gemeq}-\rfr{egos} $\bds E_g$ is the GE field, while  $\bds B_g$ is the GM one.
The resulting GM  precessions are, in GTR, \citep{LT}
\eqi
\begin{array}{lll}
\dot\Omega_{\rm LT} &=& \rp{ 2GS}{c^2 a^3 (1-e^2)^{3/2}}, \\ \\
\dot\omega_{\rm LT} &=& -\rp{6 GS\cos I}{c^2 a^3 (1-e^2)^{3/2}},
\end{array}\lb{ltprec}
\eqf
where $a,e,I$ are the semi-major axis, the eccentricity and the inclination to the equatorial plane of the central body, respectively, of the orbit of the test particle. The PPN expressions of the GM node and pericenter precessions have a multiplicative $(1+\gamma)/2$ factor
in front of $\dot\Omega_{\rm LT}$ and $\dot\omega_{\rm LT}$ \citep[p. 40]{Pet}.

In regard to the LT effect, at present, {some attempts} to measure it in the gravitational fields of the {Earth} and {Mars} have been performed with the {LAGEOS} \citep{Ciufo09} and {Mars Global Surveyor} \citep{Iorio06} artificial satellites in a series of tests exploiting non-dedicated spacecraft in an \virg{opportunistic} approach. Their status is somewhat  {uncertain}, and the {realistic evaluation of the accuracy} reached in such tests is matter of {controversy} \citep{Krogh07,Iorio09,Ciufo09,Iorio010}
The data analysis \citep{Eve09} of the {Gravity Probe B} (GP-B) mission \citep{Eve01}, aimed to {directly} measure another GM effect, i.e. the \citet{Schiff} precession of the spin of an orbiting gyroscope, in a {{dedicated}} spacecraft-based {{experiment}} orbiting the Earth, has recently released its final results \citep{Eve011}; it was proposed for the first time in 1961 \citep{Fair61}. Anyway, it will likely {{not}} be possible to {{repeat}} such an experiment in any foreseeable future because of its extreme sophistication and cost.

Concerning the Sun and its planets, it can be said that, contrary to the planet-spacecraft scenarios, the systematic errors caused by competing classical forces are of relatively less concern with respect to the measurement errors which, at present, represent the major obstacle in measuring the solar LT effect. Indeed, the non-gravitational perturbations are absent, while the gravitational ones are relatively well known and easily to model. Suffices it to say that, in contrast with, e.g., the Earth, only one even zonal harmonic coefficient $J_2$ of the multipolar expansion of the non-spherical part of its Newtonian gravitational potential has to be taken into account for the Sun.
For a long time since its prediction in {1918} within GTR, the LT precessions have been retained {too small} to be detected with {planetary motions}.
Indeed, by assuming
a homogeneous and uniformly rotating  Sun,  \citet{DeS} obtained a value of $-0.01$ arcsec cty$^{-1}$ for Mercury in his pioneering work in which he preliminarily
worked out the effects of the solar rotation on the planetary perihelia within GTR limiting to ecliptic orbits only; such a figure is also quoted by
\citet[p. 111]{Soffel}. With the same assumptions concerning the rotation of the Sun, \citet{Cug78} obtained $-0.02$ arsec cty$^{-1}$; as a consequence, all such authors concluded
that, at their time, it was not possible to measure the solar LT effect;  \citet[p. 23]{Soffel}, e.g., reports failed attempts to detect the LT effect by  fitting planetary data with modern numerically produced PPN ephemerides.
Nowadays, the expected magnitude of the LT planetary precessions is even smaller than before. It is so because recent measurements of the Sun's proper angular momentum
\eqi S_{\odot}=(190.0\pm 1.5)\times 10^{39}\ {\rm kg\ m}^2\ {\rm s}^{-1}\lb{spinz}
\eqf
from helioseismology \citep{Pij1,Pij2}, accurate to $0.8\%$, yield a value about one order of magnitude smaller than that  obtained by assuming
a homogeneous and uniformly rotating  Sun.
Despite such new findings, the perspectives of measuring the GM field of the Sun from planetary motions are more favorable now than in the past.
This can be easily recognized from the following simple arguments.
\textrm{The characteristic length with which the accuracy of the determination of the orbits of the particles should be compared}
is \eqi l^{\odot}_g \doteq \rp{S_{\odot}}{M_{\odot}c} =319\ {\rm m};\eqf
in the case of the GE effects, $l_g$ is usually replaced by the Schwarzschild radius $R_g\doteq 2GM/c^2=3$ km for the Sun.
The present-day accuracy in knowing, e.g., the inner planets' mean \textrm{orbital} radius  \eqi\left\langle r\right\rangle=a\left(1+\rp{e^2}{2}\right),\eqf is shown in
Table \ref{chebol}.
Such values have been obtained by linearly propagating the formal, statistical errors in $a$ and $e$ according to Table 3 of \citet{Pit08}; even by re-scaling them by a a factor of, say, $2-5$, the GM effects due to the Sun's rotation fall, in principle, within the measurability domain.
Another possible way to evaluate the present-day uncertainty in the planetary orbital motions consists of looking at different ephemerides of comparable accuracy.
In Table \ref{chebol} we do that for the EPM2006/EPM2008 \citep{Pit08,PROO}, and the DE414/DE421 \citep{DE414,DE421} ephemerides; although, larger than $\delta\left\langle r\right\rangle$,
the maximum differences between such ephemerides are smaller than the solar GM length $l_g^{\odot}$.

Here we will focus on the {Earth-planet range $|\vec{\rho}|$} because it is a {{direct}}, {{ unambiguous}} observable for which great improvements are expected in the near future.

The paper is organized as follows. In Section \ref{due} we deal with the numerical calculation of the impact of several dynamical perturbations on the Earth-Mercury range after having reviewed the extent of the expected future improvements in measuring it. In particular, Section \ref{tre} and Section \ref{quattro} are devoted to the general relativistic gravitoelectromagnetic fields of the Sun. In Section \ref{cinque}-Section \ref{sette} the effects of the Newtonian perturbations due to the solar oblateness, the ring of the minor asteroids, Ceres, Pallas and Vesta, and the TNOs are investigated. The conclusions are in Section \ref{otto}.
%
%
\section{The interplanetary ranging}\lb{due}
Recent years have seen increasing efforts towards the implementation of the Planetary Laser Ranging (PLR) technique accurate to cm-level \citep{ILR2,ILR1,ILR,Degn,ILR3,ILR4,MOLA,recent}. It would allow to reach major improvements in three related fields: Solar System dynamics, tests of GTR and alternative theories of gravity, and physical properties of the target planet itself. In principle, any Solar System body endowed  with a solid surface and a transparent atmosphere would be a suitable platform for a PLR system, but some targets are more accessible than others. Major efforts have been practically devoted so far to Mercury \citep{ILR1} and Mars \citep{ILR2,ILR3}, although simulations reaching 93 au or more have been undertaken as well \citep{Degn,MOLA}.
In 2005 two interplanetary laser transponder experiments were successfully
demonstrated by the Goddard Geophysical Astronomical Observatory (GGAO). The first utilized the non-optimized Mercury Laser Altimeter
(MLA) on the Messenger spacecraft \citep{ILR1,ILR}, obtaining a formal error in the laser range solution of 0.2 m, or one part in $10^{11}$. The second utilized the Mars Orbiting Laser
Altimeter (MOLA) on the Mars Global Surveyor spacecraft \citep{Abshire,ILR}. A precise measure of the Earth-Mars distance, measured between their centers of mass and taken over an extended period (five years or more), would support, among other things, a better determination of several parameters of the Solar System. Sensitivity analyses point towards measurement uncertainties between 1 mm and 100 mm \citep{ILR2}. Concerning Mercury, a recent analysis on the future BepiColombo\footnote{It is an ESA mission, including two spacecraft, one of which provided by Japan, to be put into orbit around Mercury. The launch is scheduled for 2014. The construction of the instruments is currently ongoing.} mission, aimed to accurately determining, among other things, several key parameters of post-Newtonian gravity and the solar quadrupole moment from Earth-Mercury distance data collected with  a multi-frequency radio link \citep{Milani0,Milani}, points toward a maximum uncertainty of $4.5-10$ cm in determining the Earth-Mercury range over a multi-year time span (1-8 yr) \citep{Milani0,Ashby,Milani}. A proposed spacecraft-based mission aimed to accurately measure also  the GM field of the Sun and its  $J_2$  along with other PPN parameters like $\gamma$ and $\beta$ by means of interplanetary ranging is the Astrodynamical Space Test of
Relativity using Optical Devices\footnote{Its cheaper version ASTROD I makes use of one
spacecraft in a Venus-gravity-assisted solar orbit, ranging optically with ground
stations \citep{astrod1}.} (ASTROD) \citep{astrod}.
Another space-based mission proposed to accurately test several aspects of  the gravitational interaction via interplanetary laser ranging is the Laser Astrometric Test of Relativity (LATOR) \citep{lator}; the GM field of the Sun is one of its goals.

We will concentrate on Mercury both because the magnitude of the GM signal is the largest one and in view of the aforementioned expected improvements in the accuracy in the Earth-Mercury ranging.
At present, the 1-way range residuals of Mercury from radar-ranging span 30 yr (1967-1997) and are at a few km-level (Figure B-2 of \citet{DE421}); the same holds for the 1-way Mercury radar closure residuals covering  8 yr (1989-1997, Figure B-3 a) of \citet{DE421}). There are also a pair of Mariner 10 range residuals in the 70s at Mercury at 0.2 km level (Figure B-3 b) of \citet{DE421}).

Here we outline the general approach followed.
In order to numerically obtain the effect of a given gravitational perturbation P of the Newtonian Sun's monopole on the range $|\vec{\rho}|$ between the Earth-Moon Barycenter (EMB) and a given planet,  we used MATHEMATICA to simultaneously integrate with the Runge-Kutta method the equations of motion in Cartesian coordinates of EMB and the planet considered with and without the perturbation P by using the same set of initial conditions. We adopted the ICRF/J2000.0 reference frame, with the ecliptic and mean equinox of the reference epoch, i.e. J2000, centered at the Solar System Barycenter (SSB); the initial conditions at the epoch J2000 were retrieved with the HORIZONS WEB interface by JPL, NASA. The temporal interval of the numerical integration has been taken equal to $\Delta t=2$ yr because most of the present-day available  time series of range residuals approximately cover similar temporal extensions; moreover, also the typical operational time spans forecasted for future PLR technique are similar. The basic model adopted consists of the barycentric equations of motion of the Sun, the eight planets, the Moon, Ceres, Pallas, Vesta, Pluto and Eris, to be simultaneously integrated; the forces acting on them  include the mutual Newtonian $N-$body interactions, the perturbation due to the solar quadrupolar mass moment $J_2$, the effect of two rings modeling the actions of the minor asteroids and of the Trans-Neptunian Objects (TNOs), and the post-Newtonian GEM field of the Sun  (with its general relativistic value).

In order to preliminarily evaluate the potential measurability of the effects considered, the computed differences $\Delta |\vec{\rho}|\doteq |\vec{\rho}_{\rm P}|-|\vec{\rho}_{\rm R}|$, where R refers to a reference orbit which does not contain the perturbation P, were subsequently compared to the available time series of the range residuals for the inner planets  which set the present-day accuracy level in ranging to planets \citep{DE421,PROO}. About the possibility that a given, unmodeled dynamical effect may show or not its signature in the range residuals, it must be considered that the magnitude of such an effect should roughly be one order of magnitude larger than the range residuals accuracy. This to avoid the risk that it may be absorbed and partially or totally removed from the signature in the process of estimation of the initial conditions and of the other numerous solve-for parameters in the real data reduction.

Depending on the dynamical effect one is interested in, some of the perturbations examined here are to be considered as sources of noise inducing systematic bias on the target signal. For example, if the goal of the analysis is, as in our case, the LT effect, then the range perturbation due to the TNOs is clearly a source of potential systematic error which has to be evaluated. Thus, our plots are useful to assess the level of aliasing of several potential sources of aliasing for some non-Newtonian effects and the
correlations that may occur in estimating them. Dynamical effects which are viewed as noise in a given context can also be regarded as main targets in another one; see, e.g., the proposed  determination of asteroid masses through the ASTROD mission \citep{sassi}. In this respect, the LT signature, if not properly modeled, should be regarded as a source of potential bias.
\subsection{The gravitomagnetic, Lense-Thirring field of the Sun}
\begin{figure}
\begin{center}
\includegraphics*[width=10cm,angle=0]{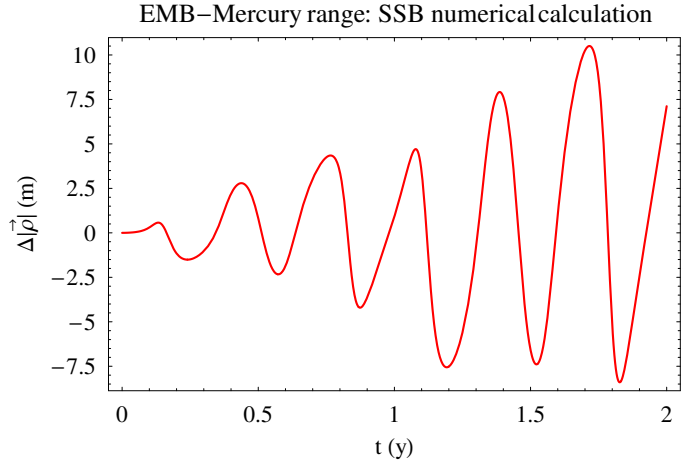}
\end{center}
\caption{Difference $\Delta |\vec{\rho}|\doteq |\vec{\rho}_{\rm P}|-|\vec{\rho}_{\rm R}|$ in the numerically integrated EMB-Mercury ranges with and without the perturbation due to the Sun's GM field over $\Delta t=2$ yr. The same initial conditions (J2000) have been used for both the integrations. The state vectors at the reference epoch have been retrieved from the NASA JPL Horizons system. The integrations have been performed in the  ICRF/J2000.0 reference frame, with  the reference $\{xy\}$ plane rotated from the mean ecliptic of the epoch to the Sun's equator, centered at the Solar System Barycenter (SSB). }\label{figz1}
\end{figure}

\lb{tre}
Figure \ref{figz1} depicts the Earth-Mercury range perturbation due to the Sun's GM field, neither considered so far  in the dynamical force models of the planetary ephemerides nor in the BepiColombo analyses.
It is important to note that the value of \rfr{spinz} for the Sun's angular momentum does not come from planetary orbital dynamics, so that there is no risk of a-priori \virg{imprinting} of  GTR itself on range tests of the solar LT effect which could, thus, be regarded as genuine and unbiased.

As a small technical note, we mention that we rotated the reference frame to the mean ecliptic at the epoch to the Sun's equator by the Carrington angle $i=7.15$ deg \citep{Carri} because \rfr{egos} holds in a frame with its $z$ axis aligned with $\bds S$.

The peak-to-peak amplitude of the LT signal  is up to $17.5$ m over 2 yr, which, if on the one hand is unmeasurable from currently available radar-ranging to Mercury, on the other hand corresponds to a potential relative accuracy in measuring it with BepiColombo of $2-5\times 10^{-3}$; this clearly
shows that the solar GM field should be taken into account in future analyses and data processing. Otherwise, it may alias the recovery of other effects, as it will become clearer later.
%
%
%
%
\subsection{The gravitoelectric, Schwarzschild field of the Sun}\lb{quattro}
In Figure \ref{EMB_Mercury_Schwa} we plot the effect of  the GE  field of the Sun on the Earth-Mercury range.
\begin{figure}
\begin{center}
\includegraphics*[width=10cm,angle=0]{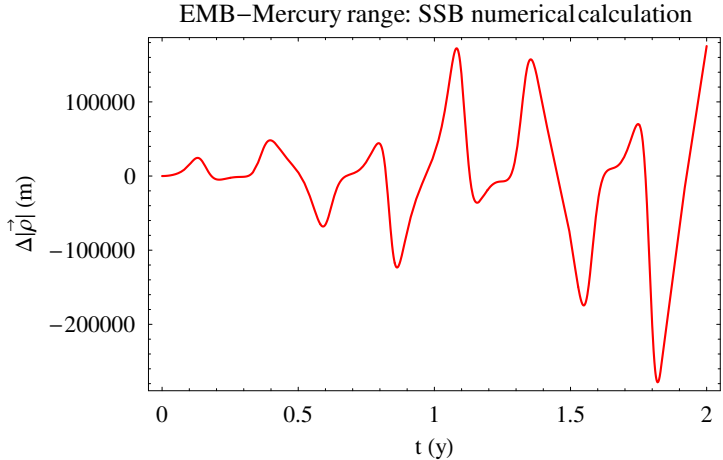}
 \end{center}
 \caption{Difference $\Delta |\vec{\rho}|\doteq |\vec{\rho}_{\rm P}|-|\vec{\rho}_{\rm R}|$ in the numerically integrated EMB-Mercury ranges with and without the perturbation due to the Sun's Schwarzschild  field over $\Delta t=2$ yr. The same initial conditions (J2000) have been used for both the integrations. The state vectors at the reference epoch have been retrieved from the NASA JPL Horizons system. The integrations have been performed in the  ICRF/J2000.0 reference frame, with the ecliptic and mean equinox of the reference epoch, centered at the Solar System Barycenter (SSB). }\lb{EMB_Mercury_Schwa}
\end{figure}

  Figure \ref{EMB_Mercury_Schwa} can be compared with Figure 1 of \citet{Milani}, obtained for unspecified initial conditions\footnote{It also includes the Shapiro delay contribution.}: they are quite similar.

  The maximum variation of the signal is of the order of $4\times 10^5$ m, corresponding to a measurement accuracy of about $2.5\times 10^{-7}$. The expected realistic accuracy in determining $\beta$ and $\gamma$ is $2\times 10^{-6}$ in BepiColombo \citep{Milani0}. To this aim, let us note that, since the LT effect depends on $\gamma$, neglecting it may alias the determination of $\gamma$ through the larger GE  signal at $4\times 10^{-5}$ level.
\subsection{The Newtonian effect of the oblateness of the Sun}\lb{cinque}
Figure \ref{EMB_Mercury_J2} shows the nominal effect of the Sun's quadrupolar mass moment  on the Earth-Mercury range for $J_2=2\times 10^{-7}$.
Its action has been modeled as \citep{VRBIK}
\eqi \bds A_{J_2}=-\rp{3J_2 R^2 GM}{2r^4}\left\{\left[1-5\left({\hat r}\cdot\hat{k}\right)^2\right]{\hat r}+2\left({\hat r}\cdot\hat{k}\right)\hat{ k}\right\}, \lb{U2}\eqf
where $R$ is the Sun's mean equatorial radius and $\hat{k}$ is the unit vector of the $z$ axis directed along the body's rotation axis.

Since \rfr{U2} holds in a frame with its $\{xy\}$ plane coinciding with the body's equator, also in this case we rotated the the mean ecliptic at the epoch to the Sun's equator.
%
%
%
%
\begin{figure}
\begin{center}
\includegraphics*[width=10cm,angle=0]{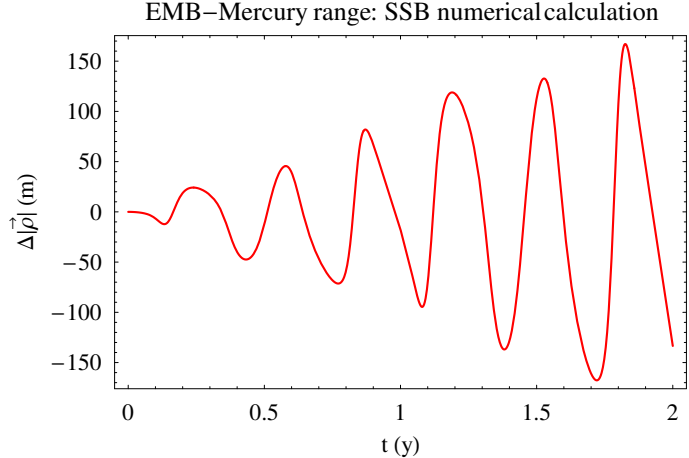}
 \end{center}
 \caption{Difference $\Delta |\vec{\rho}|\doteq |\vec{\rho}_{\rm P}|-|\vec{\rho}_{\rm R}|$ in the numerically integrated EMB-Mercury ranges with and without the nominal perturbation due to the Sun's quadrupole mass moment $J_2=2.0\times 10^{-7}$ over $\Delta t=2$ yr. The same initial conditions (J2000) have been used for both the integrations. The state vectors at the reference epoch have been retrieved from the NASA JPL Horizons system. The integrations have been performed in the  ICRF/J2000.0 reference frame, with  the reference $\{xy\}$ plane rotated from the mean ecliptic of the epoch to the Sun's equator, centered at the Solar System Barycenter (SSB). }\lb{EMB_Mercury_J2}
\end{figure}

 The signal of Figure \ref{EMB_Mercury_J2} has a maximum span of 300 m, corresponding to an accuracy measurement of $3\times 10^{-4}$. A determination of the solar $J_2$ accurate to  $10^{-2}$  is one of the goals of BepiColombo \citep{Milani0}; knowing precisely $J_2$ would yield important insights on the internal rotation of the Sun. At present, the uncertainty in it is about $10\%$ \citep{Fienga}. The solar quadrupole mass moment may play the role of source of systematic bias with respect to, e.g., some non-Newtonian dynamical effects. Concerning the GE signal previously analyzed, the mismodeled $J_2$ signature would impact it a $7.5\times 10^{-5}$ level. It is important to note that the patterns of the two signals are rather different.  Conversely,  the determination of $J_2$ at the desired level of accuracy may be affected by other unmodeled/mismdeled dynamical effects acting as systematic sources of aliasing on it.
 For example, the GM signature may affect the determination of $J_2$ at $12\%$ level. On the other hand, in order to allow for a determination of the LT effect itself,
the Sun's quadrupole mass moment should be known with an accuracy better than the present-day one by at least one order of magnitude; this is just one of the goals of BepiColombo. Anyway, also the GM and the $J_2$ patterns are different.
\subsection{The Newtonian effect of the ring of the minor asteroids and Ceres, Pallas and Vesta}\lb{sei}
In Figure \ref{EMB_Mercury_astring} we depict one potential source of systematic bias, i.e. the action of the ring of minor asteroids \citep{Fienga}.
We modeled it following \citet{Fienga08}. In general, for those planets for which
%
%
 $r<R_{\rm ring}$, by posing $\alpha\doteq r/R_{\rm ring}$, we obtained
\eqi \bds A_{\rm outer\ ring}\simeq \rp{Gm_{\rm ring}}{2rR_{\rm ring}^2}\left(\alpha+\rp{9}{8}\alpha^3+\rp{75}{64}\alpha^5\right)\bds r, \eqf
from
\eqi \bds A_{\rm outer\ ring}=\rp{Gm_{\rm ring}}{2rR_{\rm ring}^2}\left[b^{(1)}_{\rp{3}{2}}(\alpha)-\alpha b^{(0)}_{\rp{3}{2}}(\alpha)\right]\bds r.\eqf
Recall that the Laplace coefficients are defined as
\eqi b_s^{(j)}(\alpha)\doteq\rp{1}{\pi}\int_0^{2\pi}\rp{\cos j\psi d\psi}{\left(1-2\alpha\cos\psi+\alpha^2\right)^s},\eqf where $s$ is a half-integer; a useful approximate expression in terms of a series can be found in \citet[p. 237]{Murr}
\eqi
\begin{array}{lll}
b_s^{(j)}&\simeq &\rp{s(s+1)...(s+j-1)}{1\cdot3\cdot\cdot\cdot j}\alpha^j\left[1+\rp{s(s+j)}{(1+j)}\alpha^2+\right.\\ \\
&+&\left.\rp{s(s+1)(s+j)(s+j+1)}{1\cdot 2(j+1)(j+2)}\alpha^4\right].
\end{array}
\eqf
\begin{figure}
\begin{center}
\includegraphics*[width=10cm,angle=0]{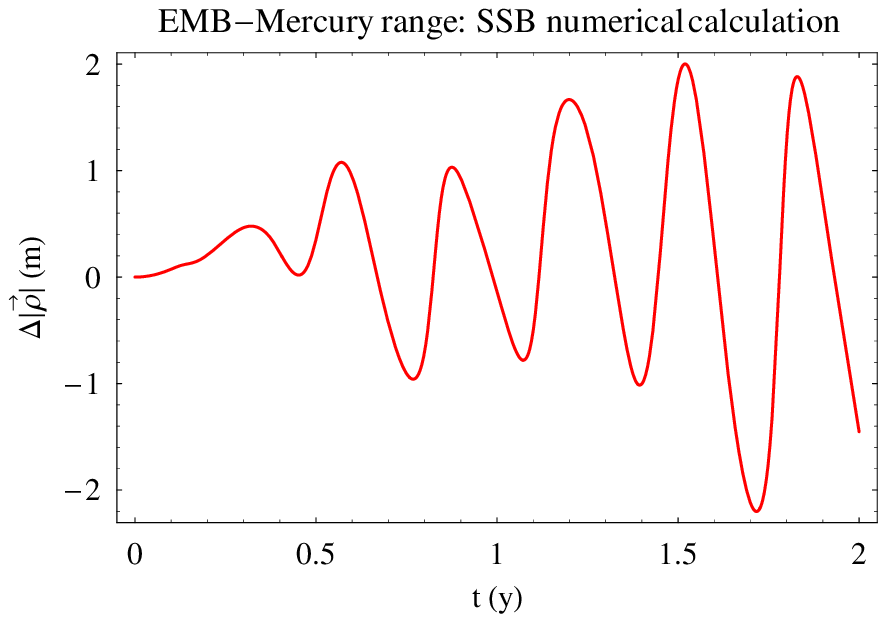}
 \end{center}
 \caption{Difference $\Delta |\vec{\rho}|\doteq |\vec{\rho}_{\rm P}|-|\vec{\rho}_{\rm R}|$ in the numerically integrated EMB-Mercury ranges with and without the nominal perturbation due to the ring of minor asteroids with $m_{\rm ring}=1\times 10^{-10}$M$_{\odot}$ \citep{Fienga} and $R_{\rm ring}=3.14$ au \citep{Fienga} over $\Delta t=2$ yr. The same initial conditions (J2000) have been used for both the integrations. The state vectors at the reference epoch have been retrieved from the NASA JPL Horizons system. The integrations have been performed in the  ICRF/J2000.0 reference frame, with the ecliptic and mean equinox of the reference epoch, centered at the Solar System Barycenter (SSB). }\lb{EMB_Mercury_astring}
\end{figure}
By assuming for the ring of the minor asteroids a nominal mass of $m_{\rm ring}=1\times 10^{-10}$M$_{\odot}$ \citep{Fienga} and a radius $R_{\rm ring}=3.14$ au \citep{Fienga},
it would impact the Mercury range at 4 m level (peak-to-peak amplitude), which will be, in fact, measurable. Its nominal bias on the Schwarzschild, $J_2$ and LT signals would be $1\times 10^{-5}, 1.3\times 10^{-2},2.3\times 10^{-1}$, respectively. Anyway, the present-day level of uncertainty in the mass of the ring is $\delta m_{\rm ring}=0.3\times 10^{-10}$M$_{\odot}$ \citep{Fienga}. Thus, the impact of such a mismodeling would be, $3\times 10^{-6}, 4\times 10^{-3}, 7\times 10^{-2}$, respectively; it cannot be considered negligible.

The effect of  Ceres, Pallas and Vesta on the determination of some Newtonian and non-Newtonian parameters with BepiColombo has been preliminarily investigated in \citet{Ashby}. Here
in Figure \ref{EMB_Mercury_CePaVe}
we
show
the nominal perturbation on the Earth-Mercury range due to the combined actions of Ceres, Pallas and Vesta; the values for their masses have been retrieved from \citet{CePaVe}.
\begin{figure}
\begin{center}
\includegraphics*[width=10cm,angle=0]{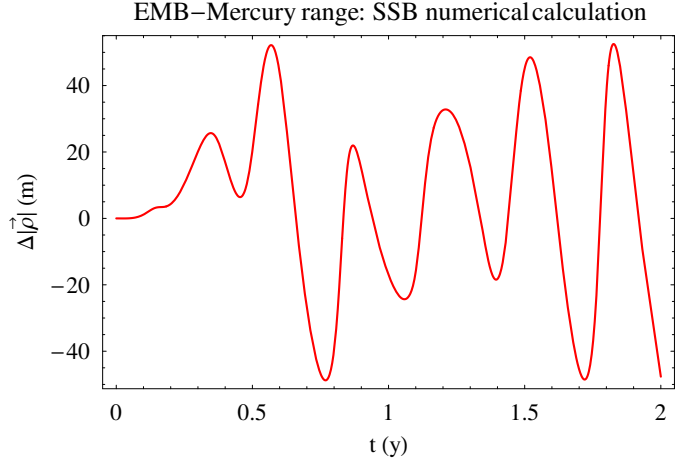}
 \end{center}
 \caption{Difference $\Delta |\vec{\rho}|\doteq |\vec{\rho}_{\rm P}|-|\vec{\rho}_{\rm R}|$ in the numerically integrated EMB-Mercury ranges with and without the nominal perturbation due to Ceres, Pallas, Vesta \citep{CePaVe} over $\Delta t=2$ yr. The same initial conditions (J2000) have been used for both the integrations. The state vectors at the reference epoch have been retrieved from the NASA JPL Horizons system. The integrations have been performed in the  ICRF/J2000.0 reference frame, with the ecliptic and mean equinox of the reference epoch, centered at the Solar System Barycenter (SSB). }\lb{EMB_Mercury_CePaVe}
\end{figure}
Its peak-to-peak amplitude amounts to 80 m; thus, their signature would be measurable at a $0.6-1\times 10^{-3}$ level. Anyway, the mismodeled solar quadrupole mass moment would bias their signal at $4\times 10^{-1}$ level. The LT effect, if unmodeled, would have an impact at $2.2\times 10^{-1}$ level.
The present-day relative uncertainties in their masses are $6\times 10^{-3}, 3\times 10^{-2}, 2\times 10^{-2}$ respectively \citep{CePaVe}. This implies a mismodeled signal with a peak-to-peak amplitude of 50 cm. It would impact the  Schwarzschild, $J_2$ and LT range perturbations at $1\times 10^{-6},2\times 10^{-3},3\times 10^{-2}$ level, respectively.
\subsection{The Newtonian effect of the Trans-Neptunian Objects}\lb{sette}
The situation is different for another potential source of systematic uncertainty, i.e. the Trans-Neptunian Objects (TNOs). Figure  \ref{EMB_Mercury_tnoring}, obtained by modeling them as a ring with $m_{\rm ring}=5.26\times 10^{-8}$M$_{\odot}$ \citep{PROO} and $R_{\rm ring}=43$ au \citep{PROO}, shows that their maximum effect would amount to 80 cm. We used the same formulas as for the asteroid ring.
\begin{figure}
\begin{center}
\includegraphics*[width=10cm,angle=0]{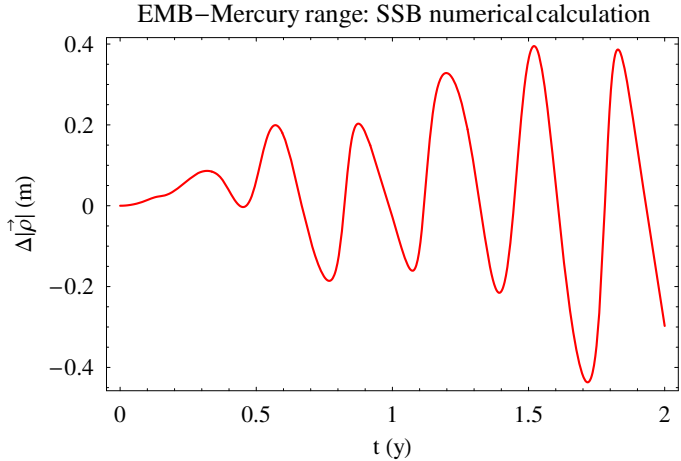}
 \end{center}
 \caption{Difference $\Delta |\vec{\rho}|\doteq |\vec{\rho}_{\rm P}|-|\vec{\rho}_{\rm R}|$ in the numerically integrated EMB-Mercury ranges with and without the nominal perturbation due to the ring of Trans-Neptunian Objects with $m_{\rm ring}=5.26\times 10^{-8}$M$_{\odot}$ \citep{PROO} and $R_{\rm ring}=43$ au \citep{PROO} over $\Delta t=2$ yr. The same initial conditions (J2000) have been used for both the integrations. The state vectors at the reference epoch have been retrieved from the NASA JPL Horizons system. The integrations have been performed in the  ICRF/J2000.0 reference frame, with the ecliptic and mean equinox of the reference epoch, centered at the Solar System Barycenter (SSB). }\lb{EMB_Mercury_tnoring}
\end{figure}
Such an effect, not taken into account so far, would be better measurable than that by the minor asteroids. This implies a bias of $2\times 10^{-6}$ on the Schwarzschild signal, $3\times 10^{-3}$ for $J_2$ and $4.5\times 10^{-2}$ for the Lense-Thirring effect. A major concern is that the mass of the TNOs is far from being accurately known, so that an uncertainty as large as $100\%$ should be applied.
\section{Summary and Conclusions}\lb{otto}
In view of a possible future implementation of some interplanetary laser ranging facilities accurate to cm-level ($4.5-10$ cm for BepiColombo in $1-8$ yr),  we have numerically investigated how the ranges between the Earth and Mercury are affected by certain Newtonian and non-Newtonian dynamical effects by simultaneously integrating the equations of motion of all the major bodies of the Solar System plus some minor bodies of it (Ceres, Pallas, Vesta) in the SSB reference frame over a time span two years long.

It turns out that the general relativistic gravitomagnetic Lense-Thirring effect of the Sun, not modeled so far either in the planetary ephemerides or in the analyses of some spacecraft-based future missions like, e.g., BepiColombo, does actually fall within the measurability domain of future  cm-level ranging facilities. The more favorable situation occurs for Mercury because the relative measurement accuracy is of the order of $2-5\times 10^{-3}$ by assuming a $4.5-10$ cm uncertainty in the Earth-Mercury ranging, as expected for BepiColombo over some years of operations.

If not properly modeled and solved-for, the Lense-Thirring effect may also impact the determination of other Newtonian and post-Newtonian parameters at a non-negligible level, given the high accuracy with which their measurement is pursued. For example, in the case of BepiColombo the expected accuracy in determining  $\gamma$ and $\beta$ from the range perturbation due to the  Schwarzschild field of the Sun is of the order of $10^{-6}$; the Lense-Thirring range perturbation would impact the Schwarzschild one at $4\times 10^{-5}$ level. Another goal of the BepiColombo mission is a  measurement of the Sun's quadrupole mass moment $J_2$ accurate to $10^{-2}$; the unmodeled Lense-Thirring effect would bias it at  $10^{-1}$ level.

From the point of view of a measurement of the Sun's gravitomagnetic field itself, it results that
a major concern would be the solar oblateness; it should be known at a $10^{-2}$ level of accuracy-which is just the goal of BepiColombo-to allow for a  reduction of its aliasing impact on  the Lense-Thirring signal down to just $17\%$.
The ring of the minor asteroids should be taken into account as well because its mismodeling would impact the gravitomagnetic signal at about $7\times 10^{-2}$. The lingering uncertainty in the masses of Ceres, Pallas, Vesta translates into a potential bias of about $4.5\times 10^{-2}-10^{-3}$. The TNOs, not modeled so far apart from the EPM ephemerides, would nominally affect it at a $4.5\times 10^{-2}$ level; it must be considered that there is currently a high uncertainty in their mass. However, it must be noted that the previous figures have been obtained by comparing the peak-to peak amplitudes of the various time-dependent range signals. Actually, the time signatures of such sources of systematic bias are different with respect to the gravitomagnetic one; this would greatly help in disentangling it from the noisy effects.

Table \ref{tavolona} summarizes our findings.

Finally, we point out that the analysis presented here should be regarded just as a preliminary investigation. Actually, extensive simulations should be performed in which a gravitomagnetic-dedicated parameter should be estimated along with the wealth of other ones which are routinely solved-for in the data reduction procedures in order to check how the relativistic signal of interest may be affected by the estimation of, say, the initial conditions of the planets/spacecraft whose data are processed. Indeed, it must be recalled that an unmodelled dynamical effect may be, partially or totally, absorbed in some of the estimated parameters, especially if its magnitude is not large enough with respect to the characteristic accuracy level of the observations. Anyway, the implementation of such a non-trivial task is beyond the scopes of the present paper, and may well constitute the subject for further researches.

\clearpage

\begin{table}
\caption{First line: uncertainties \textrm{(in m)} in the average heliocentric distances of the inner planets obtained
by propagating the formal errors in $a$ and $e$ according to Table 3 of \protect\citet{Pit08}; the EPM2006 ephemerides
were used by \protect\citet{Pit08}. Second line: maximum differences \textrm{(in m)} between the EPM2006 and the DE414
\protect\citep{DE414} ephemerides for the inner planets in the time interval 1960-2020 according to Table 5 of
\protect\citet{PROO}.
Third line: maximum differences \textrm{(in m)} between the EPM2008 \protect\citep{PROO} and the DE421
\protect\citep{DE421} ephemerides for the inner planets in the time interval 1950-2050 according to Table 5 of
\protect\citet{PROO}.
They have to be compared \textrm{with} the characteristic gravitomagnetic length of the Sun $l_g^{\odot}=319$ m.\label{chebol}
}
\begin{tabular}{lllll}
\hline
Type of orbit uncertainty & {Mercury} & {Venus} & {Earth} & {Mars}  \\
 \hline
$\delta\left\langle r \right\rangle$ (EPM2006)  & 38 & 3 & 1 & 2 \\
EPM2006$-$DE414 & 256 & 131 & 17.2 & 78.7 \\
EPM2008$-$DE421 & 185 & 4.6 & 11.9 & 233\\
 \hline
\end{tabular}
\end{table}

\begin{table}
\caption{
Maximum peak-to-peak nominal amplitudes, in m, of the Earth-Mercury range signals over $\Delta t=2$ yr due to the dynamical effects listed. We adopted the standard value $J_2^{\odot}=2.0\times 10^{-7}$ \protect{\citep{Fienga}} for the quadrupole mass moment of the Sun. It is presently known at a $10\%$ level of accuracy. For its proper angular  momentum we used $S_{\odot}=190.0\times 10^{39}$ kg m$^2$ s$^{-1}$ \protect{\citep{Pij1,Pij2}} from helioseismology. For the ring of the minor asteroids we used  $m_{\rm ring}=(1\pm 0.3)\times 10^{-10}{\rm M}_{\odot},R_{\rm ring}=3.14$ au \protect{\citep{Fienga}}, while for the TNOs, modeled as massive ring as well, we adopted $m_{\rm ring}=5.26\times 10^{-8}{\rm M}_{\odot}, R_{\rm ring}=43$ au \protect{\citep{PROO}}. The masses of the major asteroids Ceres Pallas, Vesta, accurate to $10^{-2}-10^{-3}$ level, have been retrieved from \protect{\citet{CePaVe}}.
}
\label{tavolona}
\begin{tabular}{ll}
\hline
Dynamical effect & Peak-to-peak  amplitude (m)  \\
\hline
Solar Schwarzschild & $4\times 10^5$ \\
%
Solar $J_2^{\odot}$ & 300 \\
%
Ceres, Pallas, Vesta & 80 \\
%
%
Solar Lense-Thirring & 17.5 \\
Ring of minor Asteroids & 4 \\
%
%
%
TNOs & 0.8 \\
%
%
%
%
\hline
\end{tabular}
\end{table}

\end{document}